# *In situ* observation of the generation and annealing kinetics of E' centers induced in amorphous $SiO_2$ by 4.7eV laser irradiation

F. Messina*  and M. Cannas

*Dipartimento di Scienze Fisiche ed Astronomiche dell'Università di Palermo and*

*Istituto Nazionale per la Fisica della Materia*

*Via Archirafi 36, I-90123 Palermo, Italy*

## ABSTRACT

The kinetics of E' centers induced in silica by 4.7eV laser irradiation was investigated observing *in situ* their optical absorption band at 5.8 eV. After exposure the defects decay due to reaction with diffusing molecular hydrogen of radiolytic origin. Hydrogen-related annealing is active also during exposure and competes with the photo-induced generation of the centers until a saturation is reached. The concentrations of E' and $H_2$ at saturation are proportional, so indicating that the UV-induced generation processes of the two species are correlated. These results are consistent with a model in which E' and hydrogen are generated from a common precursor Si-H.

* Corresponding author: F. Messina

Phone: +390916234218; Fax: +390916162461; E-mail: fmessina@fisica.unipa.it

Intrinsic point defects induced in silica by ultraviolet (UV) light and their thermal and temporal stability are a currently debated research field. In fact, defect conversion processes often degrade the optical transmittance of silica materials, limiting their use in several applications such as optical components and fibers.[1] One of the dominant contributions to the optical absorption (OA) of UV-irradiated silica is a band peaked at 5.8 eV associated with the paramagnetic E' center.[2,3] This defect consists in an unpaired electron in a $sp^3$ orbital of a Si atom bonded with three oxygen ($\equiv$Si•),[4] so being detectable by electron spin resonance (ESR) measurements. The stability of E' center, even at room temperature, is conditioned by its reactivity with molecular hydrogen, which is able to passivate the defect:[5-11]

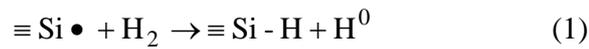

$$\equiv Si\bullet + H_2 \rightarrow \equiv Si\text{-}H + H^0 \qquad (1)$$

Reaction (1) belongs to the wide class of defect conversions in silica induced by hydrogen, quite common even well below room temperature due to the high mobility and reactivity of $H_0$ and $H_2$.[5,12] A careful study of the reaction dynamics of defects with mobile hydrogen requires spectroscopic techniques able to probe *in situ* their concentration changes. So far, photoluminescence *in situ* measurements have been used to clarify the generation and the decay of another defect of fundamental interest, the non bridging oxygen hole center (NBOHC, $\equiv$SiO•), induced in silica by photolysis of SiO-H bond.[12,13] At variance, even if a few works have investigated E' centers by monitoring optical transmittance at a fixed energy (~5.8 eV) during irradiation,[14-17] the current understanding of the generation of E' centers by UV laser is mainly founded on *ex situ* ESR and OA measurements.[18-22] Hence, the interplay between the photo-induced creation of E' center and its decay due to reaction with mobile hydrogen is not well understood. In this paper we report experimental results obtained by time-resolved *in situ* measurement of the absorption profile over the 3.1eV÷6.1eV range during and after laser exposure. This experimental approach allows to clearly determine which transient absorbing centers are induced provided that their absorption lineshapes can be unambiguously identified; then, in the present work this technique is appropriate to a comprehensive investigation of the processes controlling the generation and decay dynamics of E' centers.

Wet fused quartz, commonly employed in optical applications, is a remarkable material to study transient transmittance losses; in fact, in these glasses UV absorption due to E' centers is effective mainly during laser exposure, while transparency is almost completely recovered in the post-irradiation stage.[11] Herasil 1 samples (provided by Heraeus Quartzglas), $5\times5\times1$ mm$^3$ sized, OH content ~200 ppm, were utilized. As received this material exhibits an OA band at 5.15 eV associated with the twofold coordinated Ge impurity.[23] Moreover, the absence of the OA bands at 5.0 eV and 7.6 eV associated with the intrinsic oxygen deficient centers ODC(II) and ODC(I) respectively, [23-25] fixes an upper concentration limit of ~$10^{15}$ cm$^{-3}$ and ~$10^{16}$ cm$^{-3}$ for the two defects. Pulsed UV irradiation was performed at room temperature by a Q-switched Nd:YAG laser working in IV harmonic mode (h$\nu\approx$4.7 eV, pulse width 5 ns, repetition rate 1Hz). We performed five irradiations on different samples, with energy density varying from 13 mJ/cm$^2$ to 100 mJ/cm$^2$ per pulse. The duration of the exposures was varied from 29000 s down to 3750 up to keep constant the total incident energy density at 375 J/cm$^2$. The samples were placed with the minor surface perpendicular to the laser beam. *In situ* OA spectra were measured by an optical fiber spectrophotometer (AVANTES S2000) equipped with a D$_2$ lamp providing a ~2 $\mu$W beam on the larger surface of the sample. Transmitted light was dispersed by a grating with 1200 lines/mm and acquired by a 2048 channels charge coupled diode (CCD) array with 3ms integration time. Under irradiation, after each laser pulse we collected and averaged ten OA spectra uniformly spaced over the inter-pulse. Measurements were carried on at the same rate also for a few hours in the post-irradiation stage. OA spectra were corrected for the temporal drift of the lamp using a second reference channel. Our investigation was completed by ESR spectra performed for a few days after the end of irradiation by a Bruker EMX spectrometer working at 9.8 GHz.

Figure 1 shows the changes in OA spectra induced by 3750 laser pulses of 100 mJ cm$^{-2}$ energy density. The difference spectra measured during irradiation at increasing exposure times evidence the growth of the gaussian-shaped band centered at 5.84±0.03 eV, FWMH=0.70±0.04 eV, associated with E' centers, which reaches the maximum peak amplitude of 1.12±0.02 cm$^{-1}$. In the post-irradiation stage the band intensity progressively decreases with time, the reduction being ~50% in the first 1800 s. We note that the absorption band grows during irradiation and decreases after irradiation without shape variations, within our experimental uncertainty . Apart from the 5.8 eV band, the negative component at ~5.1eV is due to UV induced conversion of twofold coordinated Ge centers.[23,26] Furthermore, no 4.8eV band associated with NBOHC centers is observed within our experimental limit of 0.02 cm$^{-1}$, corresponding to a ~3×10$^{15}$ cm$^{-3}$ concentration on the basis of the oscillator strength (~0.05).[1,27]

The absorption cross section at 5.8eV of E' centers σ=6.4×10$^{-17}$ cm$^2$ was estimated by measuring the absorption coefficient in a reference sample in which the absolute concentration of the defects was obtained by the spin-echo decay technique.[28] Based on the fact that the absorption band of E' centers does not change in shape during time, we can use σ to calculate the concentration [E'] from the peak absorption amplitude at 5.8eV. Hence, [E'] is plotted in Fig. 2 against time. During irradiation [E'] saturates after ~2000s to a constant value [E']$_S$ = (1.7±0.1)×10$^{16}$ cm$^{-3}$. In the time instant when the laser is switched off (t=T$_{off}$) we observe a dramatic change of slope caused by the begin of E' centers annealing, as evidenced in the inset. From the initial decay slope at t=T$_{off}$ we determine the parameter Γ = [E']$_S^{-1}$ × d[E']/dt = -(1.5±0.1)×10$^{-3}$ s$^{-1}$, which measures the initial decay rate of E' centers. ESR measurements performed for a few days after irradiation show that [E'] tends to an asymptotic value lower than 10% of the maximum concentration. Analogous kinetics were observed upon irradiation with different incident power levels; the parameters [E']$_S$ and Γ are summarized in table I.

Generation of E' centers is hypothesized to occur by UV induced conversion of oxygen deficient centers (ODC), strained Si-O-Si bonds, or impurity bonds (Si-H, Si-Cl).[1,6,8,18-22] However, as above pointed out, the concentration of ODCs is too small to account for the observed values of $[E']_S$. Also the strained Si-O-Si bonds are excluded since their photolysis generates E'-NBOHC pairs,[20,22] but the latter are not observed here. Then, the most plausible precursors in the present experiment are extrinsic impurity bonds, such as Si-H and Si-Cl.

The role of single or multi-photon absorption processes in the generation mechanism of E' centers is evidenced by the power dependence of their generation efficiency;[1,6,18,22] this aspect will be dealt with elsewhere. Nevertheless, whatever the precursor and the generation mechanism may be, the finding that $[E']_S$ varies with incident power (Table I) demonstrates that the saturation during UV exposure is not ascribable to exhaustion of precursors, whose initial concentration would fix $[E']_S$ regardless the power level. Then, the saturation arises from equilibrium between two competitive processes: the photo-induced generation and the decay due to reaction with hydrogen, active also during exposure. The decay process is singled out in the post-irradiation stage when, in agreement with reaction (1) and under the stationary-state hypothesis for atomic hydrogen,[5] it is described by the following chemical rate equation:

$$\frac{1}{[E']}\frac{d[E']}{dt} = -2k[H_2] \qquad (2)$$

(the constant k is determined by the diffusion and reaction parameters of $H_2$). The decay rate $\Gamma$ corresponds to equation (2) calculated at t=$T_{off}$ ($\Gamma = -2k[H_2]_S$), $[H_2]_S$ being the available molecular hydrogen at the end of irradiation. The plot of -$\Gamma$ against $[E']_S$ (Fig. 3) evidences a linear dependence, y=αx, with α= (8.3±0.8)×10$^{-20}$ cm$^3$ s$^{-1}$, implying that $[H_2]_S$ is proportional to $[E']_S$.

As a straightforward conclusion we infer that $H_2$ available for reaction with E' is not already present in the samples before exposure, since in that case $[H_2]_S$ would be independent on irradiation, resulting in a constant value of $\Gamma$ regardless $[E']_S$. Then hydrogen has a radiolytic origin. It is known that bonded hydrogen is incorporated in silica in two main forms SiO-H and Si-H, whose UV-induced breaking can release atomic $H_0$ which dimerizes to form $H_2$.[5] The absence of NBOHC, within $\sim 3 \times 10^{15}$ cm$^{-3}$, indicates that the photolysis of SiO-H cannot account for the available $[H_2]_S$, expected to be at least one half of $[E']_S$ since E' centers decay almost completely in the post-irradiation stage. Furthermore, the proportionality between $[H_2]_S$ and $[E']_S$ leads to the hypothesis that the generation processes of the two species are not independent. In this frame, the simplest assumption is that hydrogen and E' are formed from a common precursor, i.e. the Si-H group whose dissociation produces E' centers and $H_0$ at the same amount, so that $[H_2]_S = 1/2 [E']_S$ leading to $\Gamma = -k[E']_S$, in agreement with results in Fig. 3. In this model, the best fit parameter $\alpha$ equals the reaction constant k between E' and $H_2$; this is corroborated by its close agreement with that estimated fitting the post-irradiation kinetics by Waite's theory:[11] $k = (8.4 \pm 0.5) \times 10^{-20}$ cm$^3$ s$^{-1}$.

A consequence of the precursor Si-H hypothesis is that the irradiated sample virtually returns to the same condition as the virgin material after that the recombination of E' and hydrogen in the post-irradiation stage is completed. Hence, if the sample is irradiated once more, the growth and annealing of E' centers should repeat themselves with the same kinetics observed after the first exposure. This prediction is confirmed, at least as concerns the post-irradiation stage, by a multiple-irradiations experiment recently performed on wet fused quartz, so supporting the here proposed model.[11]

In conclusion, we studied the kinetics of E' centers in amorphous $SiO_2$ exposed to 4.7eV laser light. The saturation of [E'] with time is determined by the competition between photo-generation of the defects and their simultaneous decay due to reaction with radiolytic hydrogen. Experimental results indicate that the generation processes of E' and hydrogen are correlated, consistently with a model in which the two species are formed by photo-induced breaking of a common precursor Si-H. These results prove the usefulness of *in situ* time-resolved detection of absorption spectra to perform comprehensive studies on transient point defect conversion processes and their effect on the transparency of optical materials during UV exposure.

We wish to thank our research group at University of Palermo led by Prof. R. Boscaino for support and enlightening discussions. Technical assistance by G. Lapis and G. Napoli is also acknowledged. This work is part of a national project (PRIN2002) supported by the Italian Ministry of University Research and Technology.

**FIGURE CAPTIONS**

**FIG. 1**: Difference absorption spectra measured *in situ* during (a) and after (b) a 3750 s laser irradiation with 100mJcm$^{-2}$ energy density in Herasil 1 silica sample. Arrows indicate the evolution of spectra with increasing time.

**FIG. 2**: E' centers concentration measured during and after a 3750 pulses laser irradiation at 100mJcm$^{-2}$ energy density. The inset shows a zoom of the initial decay of E' centers during the first 60 s after switching off the laser.

**FIG. 3:** Decay rate $\Gamma$ of E' centers at the beginning of the post-irradiation stage plotted against their saturation concentration [E']$_S$.

**TABLE I:** Concentration [E']$_S$ and decay rate $\Gamma$ of E' centers measured at the beginning of the post-irradiation stage at different laser energy densities.

| Energy Density (mJ cm$^{-2}$) | [E']$_S$ (cm$^{-3}$) | $\Gamma$ (s$^{-1}$) |
|---|---|---|
| 13 | $2.9 \times 10^{15}$ | $-1.0 \times 10^{-4}$ |
| 23 | $8.4 \times 10^{15}$ | $-5.4 \times 10^{-4}$ |
| 39 | $1.0 \times 10^{16}$ | $-8.7 \times 10^{-4}$ |
| 65 | $1.3 \times 10^{16}$ | $-1.2 \times 10^{-3}$ |
| 100 | $1.7 \times 10^{16}$ | $-1.4 \times 10^{-3}$ |

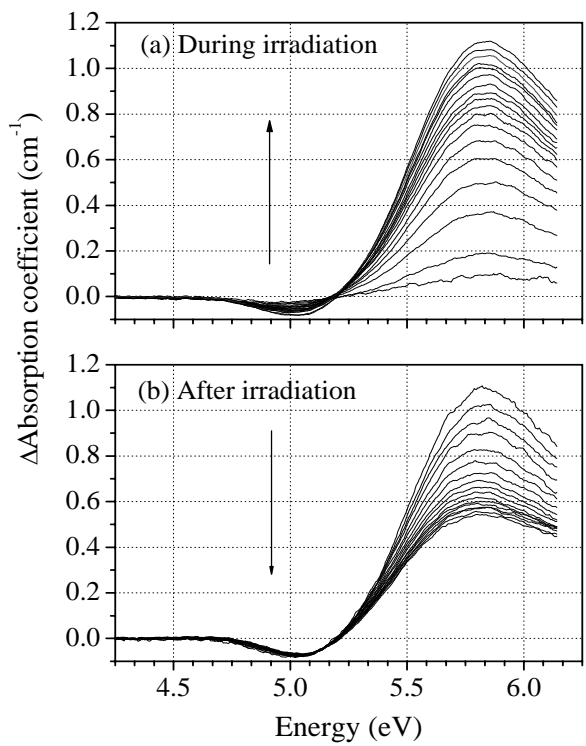

**FIGURE 1.**

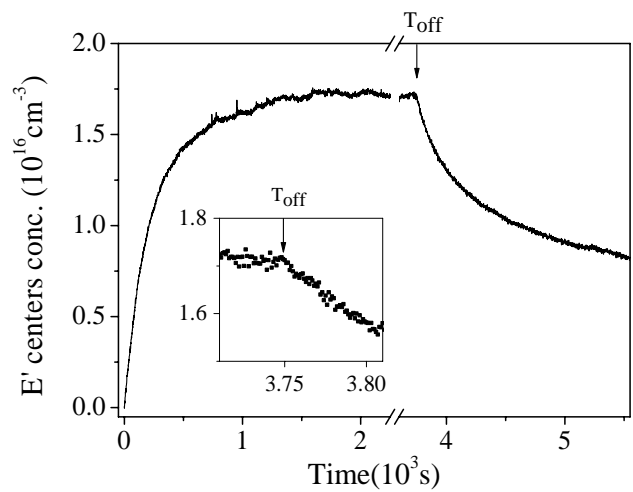

**FIGURE 2.**

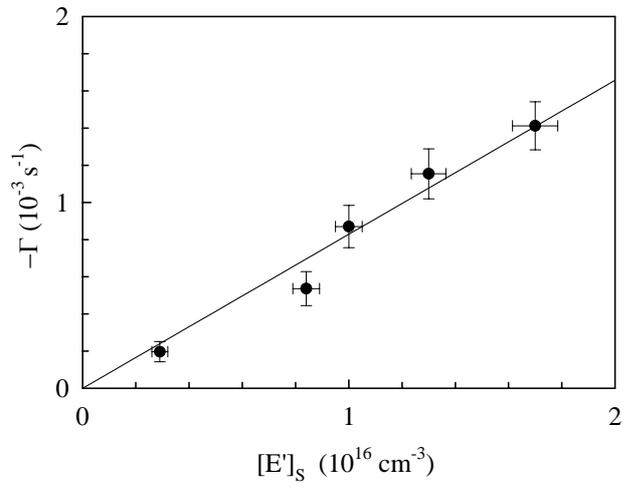

**FIGURE 3.**